\documentclass[journal, onecolumn]{IEEEtran}

\ifCLASSINFOpdf
\else
   \usepackage[dvips]{graphicx}
\fi
\usepackage{url}
\usepackage{hyperref}

\usepackage{graphicx}
\usepackage{verbatim}
\usepackage{color}

\begin{document}
\title{Demo: Pre-Characterization of Electromagnetic Side-Channel Leakage Using Publicly Available Information: A Case Study on E-Voting Interfaces}
\author{Leonardo Teodoro, Kemuel L. Vieira, Saulo Queiroz 
\thanks{This work was presented in the \emph{Show \& Tell} Technical Demonstration Session of the
\emph{IEEE International Conference on Acoustic, Speech, and Signal Processing} (ICASSP) 2026,
available in \url{https://2026.ieeeicassp.org/industry_program/\#DMOS_530}.}
\thanks{Leonardo Teodoro, Kemuel L. Vieira, and Saulo Queiroz are with the Academic Department of Informatics 
at the Federal University of Technology Paraná, Ponta Grossa 84017-220, Brazil.
(e-mail: \{lteodoro, kemuel\}@alunos.utfpr.edu.br, sauloqueiroz@utfpr.edu.br).}
}

\markboth{Show \& Tell Demo Session, IEEE International Conference on Acoustic, Speech and Signal Processsing (ICASSP), 2026.}
{Teodoro et al. \MakeLowercase{\textit{et al.}}: Pre-Characterizing Electromagnetic Leakage Using Public Information: A Case Study on Brazilian E-Voting Interfaces}
\maketitle

\begin{abstract}
In this work, we study the interface of the Brazilian e-Voting Machine (BVM) in the context of electromagnetic 
side-channel threats commonly referred to as TEMPEST attacks. In a TEMPEST attack against video displays, 
an eavesdropper uses Software-Defined Radios (SDRs) to recover sensitive information by intercepting 
electromagnetic emanations generated during video signal transmission. We emulate the BVM using a VGA 
monitor by leveraging publicly available information disclosed by the electoral authority, including 
technical specifications, operational rules of the system, and the official BVM interface. Based on this 
setup, we investigate whether the BVM interface gives rise to a distinctive spectral signature observable 
through its unintended electromagnetic emissions.
Our findings show that design characteristics relevant to a nationwide electoral process -- such as high 
image contrast, minimal on-screen information, and the prohibition of other electronic devices within the 
polling station -- result in a simple and highly distinctive spectral signature that can be observed even 
through a wall in our experiments. Although our experiments do not involve actual BVM hardware, the results 
raise concerns regarding the system's susceptibility to TEMPEST attacks and highlight the need for further 
research on protective countermeasures. In this context, our findings may support the design of automatic 
jammers capable of adaptively targeting compromising frequencies. To the best of our knowledge, this is the 
first study investigating TEMPEST attacks in the context of an electronic voting system officially adopted by a country.
\end{abstract}


\IEEEpeerreviewmaketitle

\section{Description}
Wireless side-channel attacks (SCAs) on monitor displays--often referred to 
as TEMPEST attacks--constitute a class of threats in which an eavesdropper 
remotely infers sensitive screen information by processing electromagnetic 
emanations unintentionally emitted by the display. In this demo, we present
 public TEMPEST, a variant of the TEMPEST threat model in which publicly 
available system information is leveraged to identify structural signal 
characteristics \emph{ex ante}, prior to the physical acquisition of 
electromagnetic leakage. Such pre-characterized properties can both facilitate 
subsequent side-channel exploitation and support jamming-based mitigation strategies. 
We illustrate the public TEMPEST concept through a case study based on the Brazilian 
electronic voting machine.

This research is motivated by a public call issued by the Brazilian electoral
 authority aimed at anticipating security issues in the electronic voting process 
and by a recent judicial decision that revoked a councilman’s mandate after identifying 
the use of micro-cameras to violate voting privacy. We examine how publicly available 
information about the Brazilian electoral system can expose electronic voting machines
 to TEMPEST-related SCAs.

We show that key design characteristics of the Brazilian e-voting interface--such as 
high-contrast images and minimal on-screen information adopted to improve usability for 
over 150 million electors--result in a highly distinctive spectral signature. Because these 
interfaces are publicly available, this signature can be analyzed offline and used to support 
the automatic tuning of electromagnetic parameters that vary across different e-voting machine 
models (e.g., critical harmonic frequencies).  Although this raises concerns regarding the system's 
susceptibility to TEMPEST attacks, our findings may support  the design of automatic jammers capable 
of adaptively targeting the compromising frequencies. To the best  of our knowledge, this is the first 
study in the literature about TEMPEST attacks in the context of an electronic voting system officially 
adopted by a country.

\section{The TEMPEST Signal}\label{sec:background}
In TEMPEST attacks, an eavesdropper reconstructs sensitive information from electromagnetic
 signals unintentionally emitted by the target device~\cite{tempestclass-ieeeaccess2025},~\cite{lvds-ieeetifs-25}. 
In a VGA video system, the emitted continuous-time signal $x(t)$ arises from (approximately) rectangular 
pulses $p(t - nT_p)$ associated with the pixel intensities $x[n]$:
\begin{equation}
x(t) = \sum_{n} x[n]p(t - nT_p), \label{eqn:signal}
\end{equation}
where the pulse (pixel) duration $T_p$ relates to the
pixel rate $f_p$ as follows
\begin{eqnarray}
T_p &=& \frac{1}{f_p} \quad \textrm{s}, \label{eqn:tp} \\
f_p &=& P_x P_y f_v \quad \textrm{pixels/s} \label{eqn:pixelrate},
\end{eqnarray}
and $P_x$, $P_y$, and $f_v$ 
represent the number of pixels per line (including blanking pixels), 
the number of lines per frame, and the frame rate, 
respectively. The Fourier transform of the leaked signal $x(t)$ is given by
\begin{equation}
X(f)=P(f)\sum_{n} x[n]e^{-j2\pi nfT_p},
\end{equation}
where $P(f)$ denotes the spectrum of the pulse-shaping function $p(t)$, and
$\sum_{n} x[n]e^{-j2\pi nfT_p}$
is the discrete-time Fourier transform (DTFT) of the pixel sequence $x[n]$. 
Consequently, $X(f)$ exhibits spectral replicas centered at frequencies $f=kf_p$, where $k$ is
integer and $f_p=1/T_p$. Although the ideal pulse model predicts a sinc-shaped envelope with nulls 
at several of these frequencies in theory, analog distortion of display electronics and 
propagation effects cause energy leakage around such harmonics in practice, enabling signal 
interception at suitable compromising frequencies. 

\section{Open Information as a Side-Channel Attack Enabler}\label{sec:publictempest}
In conventional TEMPEST attacks, the parameters associated with the pixel rate (\ref{eqn:pixelrate}) 
must be inferred from the eavesdropped signal (\ref{eqn:signal}) to allow proper signal 
resynchronization and image reconstruction. This can be reasonably accomplished by leveraging 
the typical correlation present in video signals (i.e., across frames, across neighbor 
pixels)~\cite{disp-eleccompa-2019}. However, public systems such as the Brazilian electronic
ballot process may, inadvertently, facilitate the accomplishment of different stages of a
TEMPEST attack by inadvertently publishing such as hardware technical specifications, 
operational rules of the system, and the official interface of the e-voting machine.
In Table~\ref{tb:publictempest}, we summarize the how some publicly available information can 
be mapped to  a distinctive spectral signature in a TEMPEST attack.

\begin{table}[h!]
\centering
\caption{Public information as enabler of TEMPEST attacks in Public E-voting Screen. \label{tb:publictempest}}
\begin{tabular}{|l|l|l|}
\hline
\begin{tabular}[c]{@{}l@{}}Open\\ Information\end{tabular} 
& \begin{tabular}[c]{@{}l@{}}Enabling\\ Aspect\end{tabular} 
& Our Case Study \\ \hline

\begin{tabular}[c]{@{}l@{}}Screen\\ Resolution\end{tabular} 
& \begin{tabular}[c]{@{}l@{}}Reduced computational\\ complexity for frame\\ rate estimation based\\ on 
decimated DFT e.g.,~\cite{sic-magazine-2025}. \end{tabular} 
& \begin{tabular}[c]{@{}l@{}}Technical specifications\\ used in public tenders\\ for manufacturers \\(models < UE2022) \end{tabular} \\ \hline

\begin{tabular}[c]{@{}l@{}}Location of\\ target device\end{tabular} 
& \begin{tabular}[c]{@{}l@{}}Identification of suitable\\ eavesdropping positions\end{tabular} 
& Polling station \\ \hline

\begin{tabular}[c]{@{}l@{}}Rules forbidding\\ operation of third-\\ party devices\end{tabular} 
& \begin{tabular}[c]{@{}l@{}}Interference-free\\ target signal\end{tabular} 
& \begin{tabular}[c]{@{}l@{}}Only e-voting machines\\ operate during an\\ polling station\end{tabular} \\ \hline

\begin{tabular}[c]{@{}l@{}}\textbf{Publication of target}\\ \textbf{interfaces for}\\ \textbf{pedagogical purposes}\end{tabular} 
& \begin{tabular}[c]{@{}l@{}}Facilitates analysis \\ of spectral signature\\ identification \end{tabular} 
& \begin{tabular}[c]{@{}l@{}}Spectral signatures can\\ be analyzed offline\\ for automatic harmonic\\ identification\end{tabular} \\ \hline

\end{tabular}
\end{table}

\section{Case Study}
In this section, we report the results of SDR-based TEMPEST attacks on a VGA
graphic system that emulates the Brazilian e-voting machine inteface. We set 
the target display to a resolution of $1280 \times 720@60$, in accordance to the 
technical specifications of the UE2020 e-voting machine model~\cite{tse_ue2020_security_specs}. 
Thus, the resulting resolution is $P_x\times P_y@f_v=1650\times 750@60$ including blank pixels~\cite{vesa_dmt_2013}.
To eavesdropp the leaked VGA signal, we employed a Ettus USRP B200 set to a sample 
rate of $f_s=54$ MS/s with a conventional digital HD TV antenna and the gr-tempest suit~\cite{grtempest-2022}. 
The target device and the radio were placed in different rooms separed by a masonry wall.

\begin{figure}[!htb]
    \caption{Strong DC component of the Brazilian e-voting user interface (left) and 
the corresponding eavesdropped ballot (right) captured from a VGA video system mimicking the Brazilian
e-voting machine under a TEMPEST attack. 
}%
    \label{fig:evotingtempest}
    \includegraphics[height = 90mm, width=185mm]{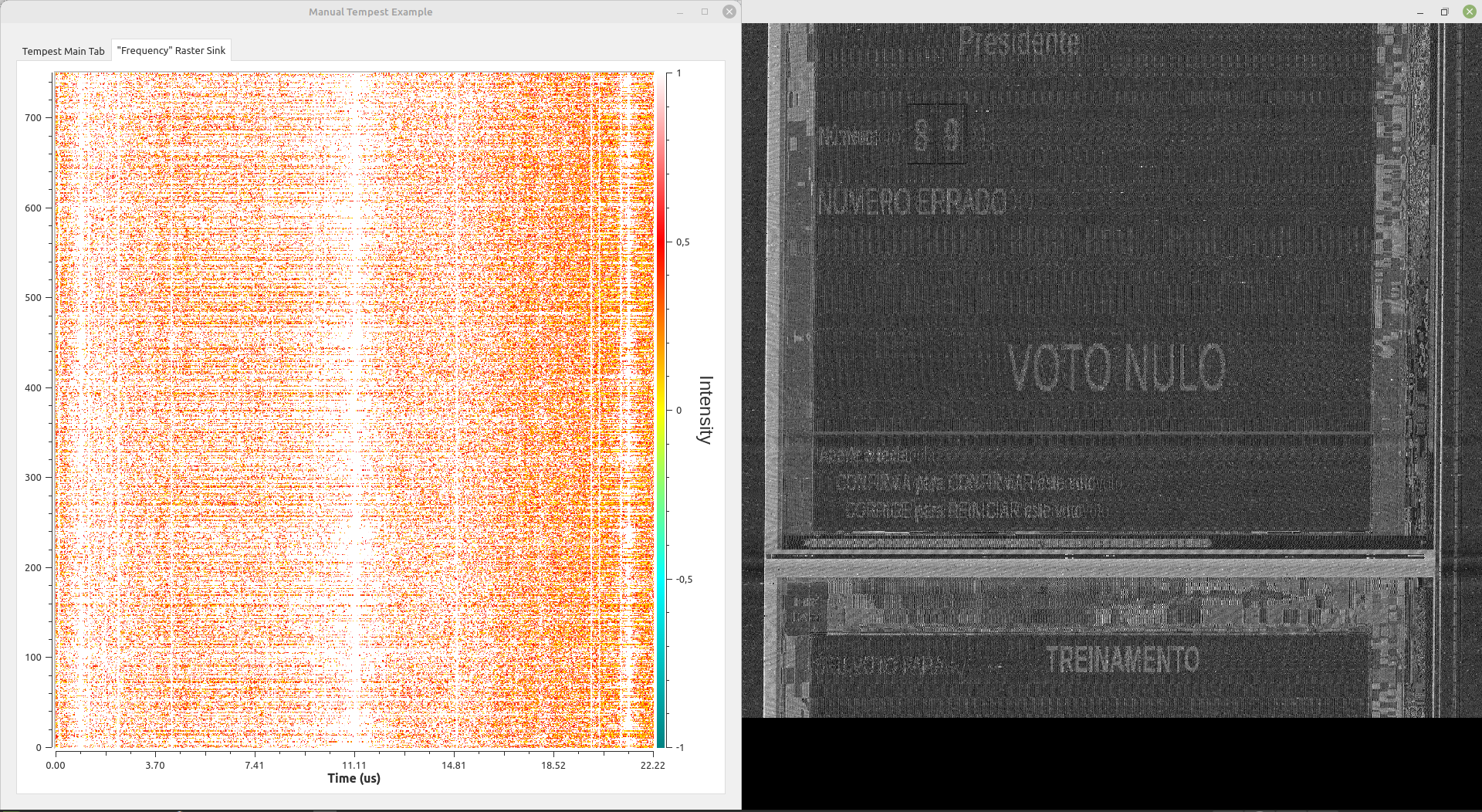}
 \end{figure}

Fig.~\ref{fig:evotingtempest} shows the GNU Radio QT GUI Time Raster graphic (left) and the
corresponding eavesdropped voting interface (right) obtained in our experiments.
The raster graphic corresponds to the line-wise DFT representation of the eavesdropped signal.
From the raster graphic, the presence of a strong component centered at the time instant 
11.11~$\mu$s is evident.
Recalling (\ref{eqn:pixelrate}) and considering $f_s=54$ MS/s, this results from the fact that only
$1650/P\times 54000000=1200$ samples, out of 1600, are acquired per line because
the SDR sampling rate operates below the pixel rate. Consequently, the line duration is
$l_s=1200/f_s\approx 22.22$ $\mu$s. Since the DC bin is shifted to the center of the graphic,
corresponding to half of the line duration, its DFT coefficient appears at the time instant
$l_s/2\approx 11.11$ $\mu$s, where the highest energy concentration is observed.
This identified pattern reveals that the high-contrast characteristic of the UE2022 interface
remains observable even after propagation through a masonry wall.

Fig.~\ref{fig:evotingtempest}, shows the GNU Radio QT GUI Time Raster graphic (left) and the
corresponding eavesdropped voting interface (right), obtained in our experiments. 
The raster graphic correspond to the line-by-line DFT of the image obtained
from the eavesdropped video signal shown on the right. From the Fig., it is clear
the presence of a very strong component centered at the time instant 11.11 $\mu$s. 
Recalling (\ref{eqn:pixelrate}) and $f_s=54$ MS/s, this stems from the fact that only 
$1650/P\times 54000000=1200$ samples, out of 1600, are acquired per line due to the 
fact that radio's sample rate operate below the pixel rate. Thus, the line duration is 
$l_s=1200/f_s\approx 22.22$ $\mu$s. Since the DC bin is shifted to the center of the graphic,
corresponding to half duration, its DFT coefficient matches the time instante 
$l_s/2\approx 11.11$ $\mu$s, where the highest energy appears. This identified pattern 
reveals that the high contrast characteristic of the UE2022 interface remains observable
even after crossing a mansory wall. 

\section{Conclusion}
In this work, we investigate whether the Brazilian e-voting machine interface gives 
rise to a distinctive spectral signature observable through its unintended electromagnetic emissions
in a TEMPEST attack. We emulate the electoral system with a VGA monitor set according to public 
available information of the Brazilian electoral system, such as, hardware technical specifications, 
operational rules of the system, and the official interface of the e-voting machine.
Our findings show that design characteristics relevant to a nationwide electoral process -- such as high 
image contrast, minimal on-screen information, and the prohibition of other electronic devices within the 
polling station -- result in a simple and highly distinctive spectral signature that can be observed even 
through a wall in our experiments. Although our experiments do not involve official e-voting hardware, the 
results raise concerns regarding the system’s susceptibility to TEMPEST attacks and highlight the need for further 
research on protective countermeasures. In this context, our findings may support the design of automatic 
jammers capable of adaptively targeting compromising frequencies. To the best of our knowledge, 
this is the first study in the literature about TEMPEST attacks against an electronic voting system 
officially adopted by a country. 

\section{Acknowledgements}
This work has been partially funded by the project Advanced Multimodal Sensing (AIMS) 
supported by Advanced Knowledge Center in Immersive Technologies (AKCIT), with financial resources 
from the PPI IoT of the MCTI grant number 057/2023, signed with EMBRAPII. The authors are also grateful 
to the ``Fundação de Amparo à Pesquisa do Estado de Goiás'' (FAPEG) for the financial support provided for this 
research (Grant 64448878/2024), the ``Secretaria de Ciência, Tecnologia e Ensino Superior (SETI)'  and  
``Fundação Araucária de Apoio ao Desenvolvimento Científico e Tecnológico do Paraná'' (call FA 17/2025).

%
\bibliographystyle{IEEEtran}
\bibliography{IEEEabrv,refs}

\end{document}